\documentclass[12pt,a4paper]{article}

\newcommand{\be}{\begin{equation}}
\newcommand{\ee}{\end{equation}}
\newcommand{\bea}{\begin{eqnarray}}
\newcommand{\eea}{\end{eqnarray}}

\begin{document}

\title{A path integral leading to higher-order Lagrangians}

\author{Ciprian Sorin Acatrinei 
\thanks{On leave from Horia Hulubei 
National Institute for Nuclear Physics and Engineering, Bucharest, MG-077125, Romania} \\
        Smoluchowski Institute of Physics, Jagellonian University \\
Reymonta 4, 30-059, Cracow, Poland \\
        {\it acatrine@th.if.uj.edu.pl}}
\date{August 27, 2007}

\maketitle

\begin{abstract}
We consider a simple modification of standard phase-space path integrals
and show that it leads in configuration space to Lagrangians 
depending also on accelerations.
\end{abstract}

Phase-space path integrals usually take the form \cite{pi}
\be
\int D\vec{q} D\vec{p}  e^{i\int_0^T dt [\vec{p} \cdot \dot{\vec{q}}-H(\vec{p},\vec{q})]}, 
\label{usual}
\ee
with border conditions enforced by the type of quantum mechanical amplitude to be evaluated.
Such integrals (or their Lagrangian counterparts) suffice for most of physical
applications, provided the symplectic structure is canonical,
$\omega_0=\sum_idp_i \wedge dq^i$.

In this note we would like to consider the following modified path integral
\be
\int D\vec{q} D\vec{p}  e^{i\int_0^T dt 
[\vec{p} \cdot \dot{\vec{q}}-
H(\vec{p},\vec{q})+\theta/2(p_1\dot{p}_2-p_2\dot{p}_1)]},
\label{current}
\ee
with $\theta$ a constant of dimension length-squared. We will 
subsequently work in two space dimensions and with all indices down,
$\vec{q}=(q_1,q_2)$, $\vec{p}=(p_1,p_2)$, for notational simplicity.
Standard notation will be used for velocity $v_i=\dot{q}_i\equiv\frac{dq_i}{dt}$,
acceleration $a_i=\ddot{q}_i\equiv\frac{d^2q_i}{dt^2}$ and mass ($m$).
The Planck constant is set to one throughout. The above apparently
innocous modification actually amounts to a change
in the symplectic structure, 
$\omega_0\rightarrow\omega=\sum_{i=1}^{2}
(dp_i \wedge dq^i+\frac{\theta}{2}dp_i\wedge dp_j)$
and has important consequences discussed below.

The path integral with modified symplectic structure (\ref{current}) happens to
describe transition amplitudes in noncommutative quantum mechanics 
\cite{b1}-\cite{a3},
but may also present further interest,
since the modification is quite simple and not too unnatural
(it is a sort of magnetic field, but in momentum space \cite{a3}).
More precisely  (\ref{current}) describes quantum mechanics 
with an additional nonvanishing commutator between coordinates, $[q_1,q_2]=i\theta$.
This  theory admits a first principles path integral formulation only in phase space,  as detailed in \cite{a2}.
At the classical level, the extended symplectic structure features 
an additional nonzero Poisson bracket, $\{q_1,q_2\}=\theta\neq 0$, 
and the resulting equations of motion do not admit a 
standard Lagrangian formulation \cite{a3}.

Nevertheless one may enforce an (effective) Lagrangian formulation in configuration space
by integrating over the momenta in the path integral (\ref{current}).
This process is described here. We first perform the calculation 
and then discuss the result.

{\bf Path integral}

We path-integrate over the momenta in (\ref{current}), to obtain the effective Lagrangian.
Starting from the partition function
\be
\int Dq_1 Dq_2 Dp_1 Dp_2 e^{i S}
\ee
with action
\be
S=\int_0^T dt
[p_1\dot{q}_1+p_2\dot{q}_2+\frac{\theta}{2}(p_1\dot{p}_2-p_2\dot{p}_1)
-\frac{p_1^2}{2m}-\frac{p_2^2}{2m}-V(q)],
\label{S_cl}
\ee
we wish to integrate over the momenta $p_1$, $p_2$. 
The potential part $V(q)$ depends only on $q_1$ and $q_2$ and plays no role
in what follows (the method is valid for any $V(q)$, 
more precisely for any Hamiltonian with separate
quadratic dependence upon momenta). We divide
the time interval $T$ in $n$ subintervals $\epsilon=\frac{T}{n}$ 
($n\rightarrow\infty$ achieves the continuum limit),
and choose for simplicity the discrete derivative 
$v^{(k)}\equiv \dot{x}^{(k)}\equiv \frac{x^{(k+1)}-x^{(k)}}{\epsilon}$;
no issues requiring symmetric operations of any kind appear in the following.
The relevant part of the discretized action (excluding $V(q)$ for now) becomes
\be 
\tilde{S}=\sum_{k=0}^{n} 
\left[
\epsilon p_1^{(k)}v^{(k)}_1+\epsilon p_2^{(k)}v^{(k)}_2+
\frac{\theta}{2}(p_1^{(k)}p_2^{(k+1)}-p_2^{(k)}p_1^{(k+1)})
-\epsilon\frac{(p_1^{(k)})^2+(p_2^{(k)})^2}{2m}
\right].
\ee
The clearest way to proceed with the coupled Gaussian integrals is to introduce matrix notation.
Define the column vectors 
\be
V\equiv \epsilon(v_1^{(0)},v_1^{(1)},\dots,v_1^{(n)}\dots ,v_2^{(0)},v_2^{(1)},\dots,v_2^{(n)}\dots)^T
\ee
\be
P\equiv (p_1^{(0)},p_1^{(1)},\dots,p_1^{(n)}\dots ,p_2^{(0)},p_2^{(1)},\dots,p_2^{(n)}\dots)^T
\ee
and the matrix
\begin{displaymath}
J = -a\left ( 
\begin{array}{ccccccccccccc}
1 & 0 & 0 & \cdot & \cdot &  &  0 & b & 0 & \cdot & \cdot &  \\
0 & 1 & 0 & \cdot & \cdot &  &  0 & 0 & b & \cdot & \cdot &   \\
\cdot & \cdot & \cdot & \cdot &  &  & \cdot & \cdot & \cdot & \cdot &  &  \\
0 & -b & 0  & \cdot & \cdot &  &  1 & 0 & 0 & \cdot & \cdot &  \\
0 & 0 & -b  & \cdot & \cdot &  &  0 & 1 & 0 & \cdot & \cdot &  \\
\cdot & \cdot & \cdot & \cdot &  &  & \cdot & \cdot & \cdot & \cdot &  &    
\end{array}               
\right ).
\end{displaymath}
where $a=\frac{\epsilon}{2m}$, $b=\frac{m\theta}{\epsilon}$. Its inverse $J^{-1}$
has the same form as above, but with different entries $a'$, $b'$, namely
$a'=1/a$ and $b'=-b$ (the off diagonal part changes sign and the overall factor
is reversed). In matrix notation the discrete action becomes
\be
\tilde{S}=P^T V + P^T J P.
\ee
The coordinate transformation 
\be 
\bar{P}\equiv P+\frac{1}{2}J^{-1} V
\ee
does not change the path integral measure ($D\bar{P}=DP$), and leads to 
\be
\tilde{S}= \bar{P}^T J \bar{P}-\frac{1}{4} V^T J^{-1} V.
\ee 
The first term is now integrated out - and no more dependency upon momenta appears, 
whereas the second term leads to an exponent of the form (modulo a factor of $i$)
\be
-\frac{1}{4} V^T J^{-1} V=\sum_{k=0}^{n}
[\epsilon \frac{m}{2}(v_1^{(k)})^2+\epsilon \frac{m}{2}(v_2^{(k)})^2
-\frac{\theta m^2}{2}(v_1^{(k)}v_2^{(k+1)}-v_2^{(k)}v_1^{(k+1)})].
\ee
Upon taking the continuum limit $\epsilon\rightarrow 0$  our main result follows:
\be
\int Dq_1 Dq_2 Dp_1 Dp_2 e^{i S}=N\int Dq_1 Dq_2 e^{i\int_0^T dt L_{eff}(q_i,v_i,a_i)}
\ee
with
\be
L_{eff}=\frac{m}{2}(\dot{q}_1^2+\dot{q}_2^2)-
\frac{\theta m^2}{2}(\dot{q_1}\ddot{q_2}-\dot{q_2}\ddot{q_1})-V(q_1,q_2)
\label{L_eff}
\ee
and $N$ a constant not depending on the $q$'s.
We have reintroduced the potential term, which passed unscathed through
Eqs. (\ref{S_cl}) -- (\ref{L_eff}). 
The second term in (\ref{L_eff}) is the correction due to noncommutativity;
it depends on velocities {\it and} accelerations,
and has an universal character. 
Its relative simplicity is striking and somehow unexpected.
One is reconforted to
find that the (initially $V=0$ version of the) 
Lagrangian (\ref{L_eff}) was studied by Lukierski et al. \cite{b21}
and shown to engender a noncommutative structure. 
A more detailed discussion follows.

{\bf Discussion}

As already mentioned, cf. \cite{a2,a3},
the resulting effective Lagrangian could not be a standard one,
depending only on coordinates and velocities. Given the complications introduced by noncommutativity, one may have expected {\it a priori} an involved function, perhaps nonlocal
or potential-dependent.
Remarkably, the effective Lagrangian turned out to be
the usual one, plus an universal correction depending also on the particle accelerations,
\be
\Delta L= -\frac{1}{2}\theta m^2 (v_1 a_2-v_2 a_1), \label{correction}
\ee
$\theta$ denoting the noncommutative scale, $m, v_i, a_i$ the mass,
velocity, respectively acceleration along the i-axis, of a given particle

Exactly the term  (\ref{correction}) was previously studied in detail in \cite{b21},
although its appearance can be traced back to earlier developments (cf. \cite{b31,b51,b42}). 
Lukierski et al. \cite{b21} started from considerations of Galilean invariance in (2+1)-dimensions,
and added  (\ref{correction}) to a free Lagrangian $\frac{m}{2}\vec{v}^2$, to provide
a dynamical realization for a free particle Galilean algebra with one extra central charge.
Upon constrained quantization of this higher order action 
(which thus circumvents the no-go theorem of \cite{a3}) 
noncommutative dynamics was shown to emerge for appropriate choices of canonical variables.
Two negative-energy "internal modes"
were proved harmless since they decoupled from the four relevant degrees of freedom.
Interactions were subsequently introduced in a constrained way in order to keep the ghosts harmless, and were described by potentials depending
on noncommutative coordinates.

In this note we went (with a different motivation) the opposite way,
starting from a Hamiltonian path integral describing
arbitrary systems with Heisenberg noncommutativity of coordinates.
To our knowledge a direct  derivation of a higher order Lagrangian from the 
extended Hamiltonian formalism was never presented before. 
The inverse - Lagrangian to Hamiltonian -
analysis of \cite{b21} indeed suggests (\ref{correction}) as an interesting possibility
(as already pointed in \cite{b31}) but does not single it out.
The maximal order of the derivatives appearing in the effective Lagrangian is not fixed apriori. 
Thus a univoque derivation was desirable. It was provided here 
using path integral methods. We obtained the 
additional acceleration-dependent term of \cite{b21}, up to coefficients,
and such correction turned out (somehow surprisingly) to be the
{\it only} possibility available for noncommutative systems of Heisenberg type and
Hamiltonians of the form 
$H=\frac{1}{2m}(p_1^2+p_2^2)+V(q_1,q_2)$.

The price to be paid for the initial noncommutativity of the coordinates
is the appearance of second order time derivatives in the action, 
and the ensuing lack of appropriate
boundary/initial conditions for the two irelevant ghost-like additional degrees of freedom.
Indeed, the classical equations of motion 
engendered by (\ref{L_eff}) are of third order in time derivatives,
\be
\epsilon_{ij} \theta m^2 \frac{d^3 q_j}{dt^3}+m\ddot{q}_i+\partial_{q_i}V=0.
\ee
No fourth-order time derivatives arise for $q_1$, $q_2$,
and this leads to two constraints in the Hamiltonian formulation.
Six constants are required - two more in comparison with the commutative case;
only four are available (for instance the initial and final values of $q_1$ and $p_2$).
This apparent indeterminacy is a consequence of the initial noncommutativity of $q_1$ and $q_2$,
but poses no serious problem. The missing two constants are actually needed to specify
the motion of the two "internal" modes, modes which 
must be eliminated for consistency, cf. \cite{b21} (see also \cite{b42}).

In conclusion, we showed that {\it all}  systems with noncommutative coordinates and
Hamiltonians of the form $p^2+V(q)$ can be described in configuration space via 
relatively simple higher-order Lagrangians.
We went in opposite direction with respect to  Lukierski et al., though 
with quite different methods and ideology. We used path integration;
no obvious reciprocal of the canonical approach of \cite{b21} is known to us at present.
Our derivation started {\it ab initio} with arbitrary potentials $V(q_1,q_2)$, 
in contrast to the inverse route taken in Ref. \cite{b21},
where  the (in the end noncommuting) variables were first 
carefuly pinned down in the free theory.

{\bf Acknowledgements}

The author was supported by
the EU Marie Curie Host Fellowships for Transfer of Knowledge Project COCOS
(contract MTKD-CT-2004-517186)
and by
NATO Grant PST.EAP.RIG.981202.


\end{document}